\documentclass[utf8]{frontiersSCNS} % for Science, Engineering and Humanities and Social Sciences articles
\usepackage{listings}
\usepackage{algorithmic}
\usepackage{graphicx}
\usepackage{caption}
\usepackage{subcaption}
\usepackage{url}
\usepackage{url,hyperref,lineno,microtype,subcaption}
\usepackage[onehalfspacing]{setspace}

\def\firstAuthorLast{Vibhatha Abeykoon et al} 
\def\Authors{Vibhatha Abeykoon\,$^{3,*}$, 
Supun Kamburugamuve\,$^{1}$ 
Chathura Widanage\,$^{1}$ 
Niranda Perera\,$^{1}$ 
Ahmet Uyar\,$^{1}$
Thejaka Amila Kanewala\,$^{3}$ 
Gregor von Laszewski\,$^{1}$ and 
Geoffrey Fox\,$^{2}$}
\def\Address{$^{1}$Luddy School of Informatics, Computing and Engineering, Bloomington, IN 47408, USA\\
$^{2}$Biocomplexity Institute \& Initiative and Computer Science Dept., University of Virginia
\\$^{3}$ Indiana University Alumni, IN 47408, USA}
\def\corrAuthor{Indiana University Bloomington}

\def\corrEmail{vibhatha@gmail.com, skamburu@iu.edu, cdwidana@iu.edu, dnperera@iu.edu, auyar@iu.edu, thejaka.amila@gmail.com, laszewski@gmail.com, gcfexchange@gmail.com}

\newcommand\archname{\textit{HPTMT}}

\begin{document}
\onecolumn
\firstpage{1}

\title[\archname{} Applications]{\archname{} Parallel Operators for High Performance Data Science \& Data Engineering} 

\author[\firstAuthorLast ]{\Authors} %This field will be automatically populated
\address{} %This field will be automatically populated
\correspondance{} %This field will be automatically populated

\extraAuth{}% If there are more than 1 corresponding author, comment this line and uncomment the next one.
%\extraAuth{corresponding Author2 \\ Laboratory X2, Institute X2, Department X2, Organization X2, Street X2, City X2 , State XX2 (only USA, Canada and Australia), Zip Code2, X2 Country X2, email2@uni2.edu}

% https://www.overleaf.com/project/60df5b982f1f0cd109e2296b
\maketitle

\begin{abstract}

% Data-intensive applications impact many domains, and their steadily increasing size and complexity demands high-performance, highly usable environments. We integrate a set of ideas developed in various data science and data engineering frameworks. They employ a set of operators on  specific data abstractions that include vectors, matrices, tensors, graphs, and tables. Our key concepts are inspired from systems like MPI, HPF (High-Performance Fortran), NumPy, Pandas, Spark, Modin, PyTorch, TensorFlow, RAPIDS(NVIDIA), and OneAPI (Intel). Further, it is crucial to support different languages in everyday use in the Big Data arena, including Python, R, C++, and Java. We note the importance of Apache Arrow and Parquet for enabling language agnostic high performance and interoperability. In this paper, we propose \textit{High-Performance Tensors, Matrices and Tables} (\archname{}), an operator-based architecture for data-intensive applications, and identify the fundamental principles needed for performance and usability success. We illustrate these principles by a discussion of examples using our software environments, Cylon and Twister2 that embody \archname{}.

Data-intensive applications are becoming commonplace in all science disciplines. They are comprised of a rich set of sub-domains such as data engineering, deep learning, and machine learning.  These applications are built around efficient data abstractions and operators that suit the applications of different domains. Often lack of a clear definition of data structures and operators in the field has led to other implementations that do not work well together. The \archname{} architecture that we proposed recently, identifies a set of data structures, operators, and an execution model for creating rich data applications that links all aspects of data engineering and data science together efficiently. This paper elaborates and illustrates this architecture using an end-to-end application with deep learning and data engineering parts working together. 

\end{abstract}

% \tiny
%  \keyFont{ \section{Keywords:} 
%  Data intensive applications, 
%  Operators, 
%  Vectors, 
%  Matrices, 
%  Tensors, 
%  Graphs, 
%  Tables, 
%  DataFrames}

\section{Introduction}

\begin{figure}[htpb]
\begin{center}
\includegraphics[width=0.7\textwidth]{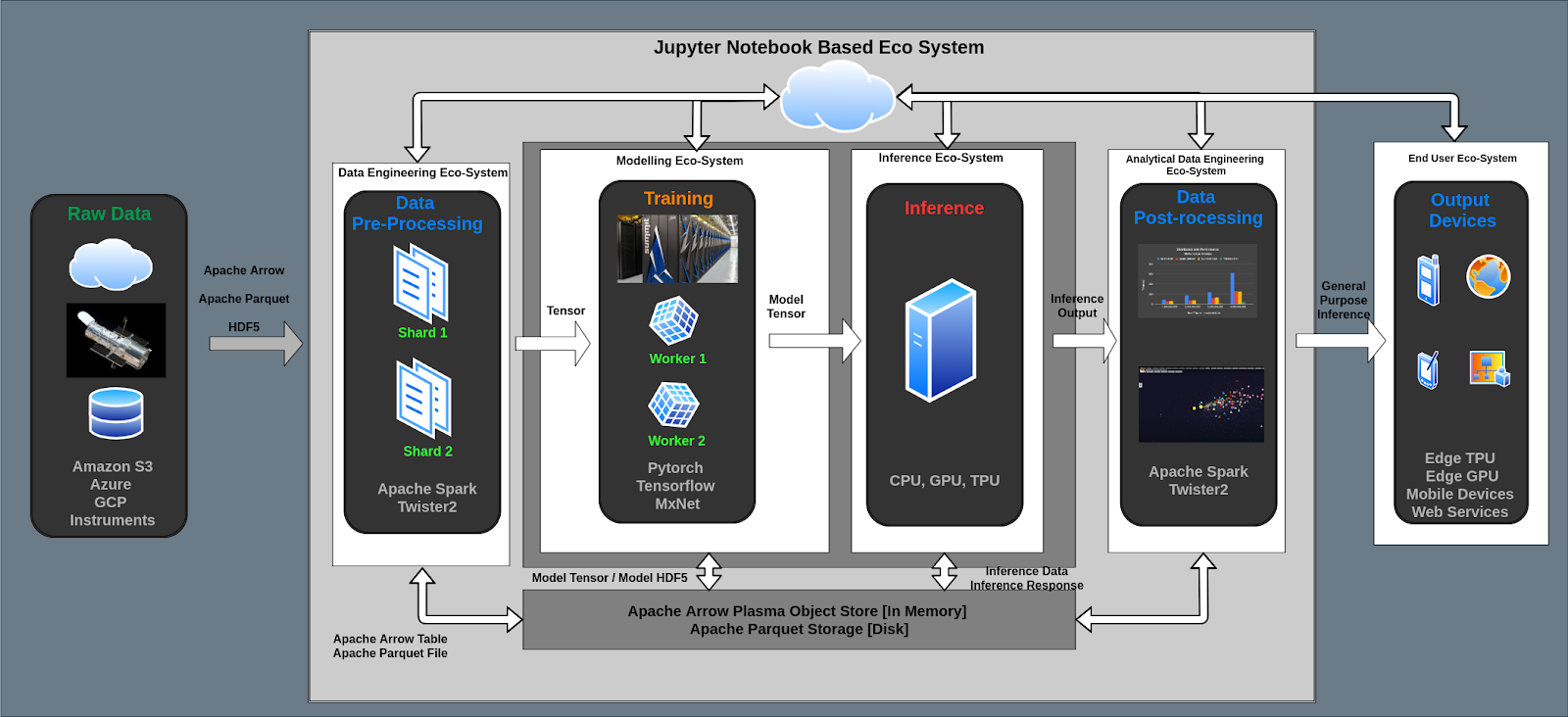}
\end{center}
\caption{Data science workflow with Jupyter Notebook interface and Data Engineering around Deep Learning}
\label{fig:datascience-workflow}
\end{figure}

Data engineering and data science are two major branches of data-intensive applications. Data engineering deals with collecting, storing, and transforming data. Data science tasks are mainly machine learning and deep learning, where we use data to learn and gain insights. These two components, illustrated in Figure~\ref{fig:datascience-workflow} are designed on top of data structures and operators around them. The data engineering component primarily works with table data abstractions, while the machine learning and deep learning components mainly use tensors and matrices. 

To run applications using multiple computers, we can partition the data and apply distributed operators. Any big data framework must uniformly support distributed data structures. Current systems use several different strategies to provide distributed application programming interfaces (APIs) for data-intensive applications. An API for data-intensive applications is a combination of data structures, operators, and an execution model. There are thousands of operators defined around data structures such as vectors and tables by different frameworks. The current data systems use asynchronous and loosely synchronous execution models for running programs at scale. Asynchronous execution is popular in systems such as Spark~\citep{spark2010}, Dask~\citep{rocklin2015dask} and Modin~\citep{petersohn2020towards}. Loosely synchronous distributed execution is used in systems such as PyTorch~\citep{pytorch}, Cylon~\citep{widanage2020high} and Twister2~\citep{twister2}. 

\begin{figure}[htpb]
\begin{center}
\includegraphics[width=0.7\textwidth]{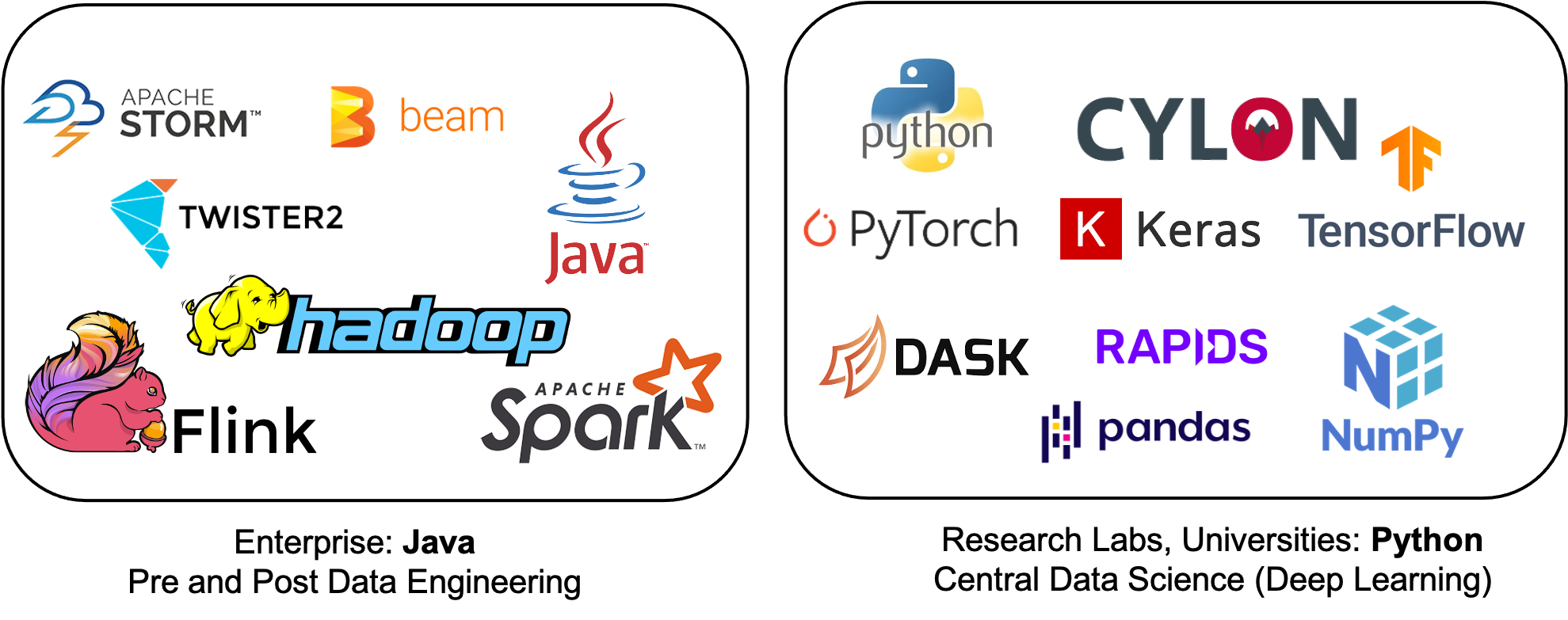}
\end{center}
\caption{The goal of HPTMT to achieve High Performance in each ecosystem and high-performance integration between ecosystems}
\label{fig:ecosystems}
\end{figure}

In a previous paper~\citep{kamburugamuve2021hptmt}, the authors proposed the \archname{} (High-Performance Tensors, Matrices, and Tables); an operator-based architecture for data-intensive applications as a scalable and interoperable way to designing rich data-intensive applications. With \archname{} we focus, as depicted in Figure~\ref{fig:ecosystems}, on the interoperability of distributed operators and how one can build large-scale applications using different data abstractions. This paper will showcase the importance of this architecture through an application that uses various data abstractions in a single distributed environment to compose a rich application. It highlights the scalability of the architecture and its applicability to high-performance computing systems. 

The rest of the paper is organized as follows. Section ~\ref{hptmt} gives an overview of the ~\archname{} architecture. Section~\ref{frameworks} describes the distributed execution of various frameworks and how they can work together according to the ~\archname{}. Section~\ref{unomt} describes an end-to-end application while section~\ref{s:s:performance-evaluation} highlights the performance. In section~\ref{related} we describe related work and conclude in section~\ref{conclude}.

\section{\archname{} Architecture}
\label{hptmt}

\archname{} architecture defines an operator model along with an execution model for scaling data-intensive applications. The primary goal of \archname{} is the efficient composability of distributed operators around different data structures to define rich applications. We see this architecture as a good candidate for exascale software environments. Its simple premise -- put the parallelism into interoperable libraries seems practical to implement well on heterogeneous collections of accelerators and CPUs. Note that one of the most successful approaches to parallel computing is based on the use of runtime libraries of well-implemented parallel operations. This was for example a key part of High-Performance Fortran HPF~\citep{dongarra2003sourcebook} and related parallel environments (HPJava~\citep{carpenter1998hpjava}, HPC++~\citep{johnson1997hpc}, Chapel~\citep{chamberlain2007parallel}, Fortress~\citep{allen2005fortress}, X10~\citep{charles2005x10}, Habanero-Java~\citep{imam2014habanero}). Such systems had limited success, maybe because the HPC community did not define sufficient operators to cover the sophisticated computational science simulations largely targeted by those languages with typically sparse or dense matrix operators. However, data-intensive applications have used similar ideas with striking success. 
\newline

\subsection{Operators}

An application domain such as deep learning or data engineering comprises a slew of combined operators to build the total job. Based on the data distribution, these operators can be categorized into two groups, namely, local operators (single machine) and distributed operators (across multiple machines). Some operators are purely local or purely distributed, and some can be either. A local operator only works with a single piece of data inside the memory of a single node in a cluster. They give rise to what is called embarrassingly or pleasingly parallel models for distributed execution. Operator based methods are not just used to support parallelism but have several other valuable capabilities

\begin{itemize}
    \item Allow interpreted languages to be efficient as overhead is amortized over the execution of a (typically large) operation
    \item Support mixed language environments where invoking language (e.g. Python) is distinct from the language that implements the operator (e.g. C++)
    \item Support proxy models where user programs in an environment that runs not just in a different language but also on a different computing system from the executing operators. This includes the important case where the execution system includes GPUs and other accelerators.
    \item Support excellent performance even in non-parallel environments. This is the case for Numpy and Pandas operators.
\end{itemize}

Recently Apache Arrow~\citep{apache-arrow} and Parquet~\citep{apache-parquet} provide essential tools which are crucial to our approach to \archname{}, and they or equivalent technologies are vital for any high-performance multi-language multi-operator class system. They provide efficient language-agnostic column storage for Tables and Tensors that allows vectorization for efficiency and performance. Note that distributed parallel computing performance is typically achieved by decomposing the rows of a table across multiple processors. Then within a processor, columns can be vectorized. This, of course, requires a large amount of data so that each processor has a big enough workflow to process efficiently. It is a well-established principle that the problem needs to be large enough for the success of parallel computing~\citep{fox2014parallel}, which the latest Big Data trends also follow. Note that the most compelling parallel algorithms use block (i.e. row and column) decompositions in scientific computing to minimize communication/compute ratios. Such block decompositions can be used in Big Data~\citep{huai2014major} (i.e. table data structures), but could be less natural due to the heterogeneous data within it.

For Big Data problems, individual operators are sufficiently computationally intensive to consider the basic job components as parallel operator invocations. Any given problem typically involves the composition of multiple operators into an analytics pipeline or more complex topology. Each node of the workflow may run in parallel. This can be efficiently and elegantly implemented using workflow such as Parsl~\citep{babuji2018parsl}, Swift~\citep{wilde2011swift}, Pegasus~\citep{deelman2015pegasus}, Argo~\citep{argo}, Kubeflow~\citep{kubeflow}, Kubernetes~\citep{kubernetes} or dataflow (Spark, Flink, Twister2) preserving the parallelism of \archname{}.
\newline

\subsubsection{Categorizing Operators}

There are thousands of operators defined for arrays, tensors, tables, and matrices. Note that tensors are similar to arrays but have an important deep learning utility. Matrices are similar to arrays and tensors but typically two dimensional. Tables including dataframes are characterized by entries of heterogeneous types. This is clear from databases where the different columns can have strings to dates to numbers. Table~\ref{tab:tensor_ops} shows some common operator categories for tensors as defined by PyTorch, Tensorflow or Keras. These deep learning frameworks define over 700 operators on tensors. Numpy lists 1085 array operations. Table~\ref{tab:table_aux_op} shows some of the popular operations on tables, where the Python Pandas library has around 224 dataframe operators out of a listed total of 4782. Also, optimized linear algebra operators are used internally in most widely used math and tensor compute libraries. Table \ref{tab:blass_op} contains a classification of BLAS operators, which are local or distributed. The (old but standard) library SCALAPACK has 320 functions (operators) at a given precision and a total of over one thousand.

\begin{table}[htpb]
\caption{Sample set of tensor operations as specified by PyTorch} 
\label{tab:tensor_ops}
\centering
\begin{tabular}{|p{3cm}|p{13cm}|}
\hline
\textbf{Operation Class} & \textbf{Description}                                         \\ \hline
Create          & Create tensors from files, in-memory data or other data structures such as NumPy \\ \hline
Math &  Multiplication, addition \\ \hline
Statistics & Statistical function such as mean, median, std \\ \hline
Indexing & Different methods to access values of tensors \\ \hline
Conversion   & Convert a tensor to another format such as NumPy or change the shape of a tensor  \\ \hline                 
\end{tabular}
\end{table}

\begin{table}[htpb]
\caption{Operators on Tables} 
\label{tab:table_aux_op}
\centering
\begin{tabular}{|p{3cm}|p{13cm}|}
\hline
\textbf{Operator} & \textbf{Description}                                                                                               \\ \hline
Select            & Filters out some records based on the value of one or more columns. \\ \hline
Project           & Creates a different view of the table by dropping some of the columns. \\ \hline
Union             & Applicable on two tables having similar schemas to keep all the records from both tables and remove the duplicates. \\ \hline
Cartesian Product & Applicable on two tables having similar schemas to keep the set of all possible record pairs that are present in both tables.       \\ \hline
Difference        & Retains all the records of the first table, while removing the matching records present in the second table.    \\ \hline
Intersect         & Applicable on two tables having similar schemas to keep only the records that are present in both tables.          \\ \hline
Join              & Combines two tables based on the values of columns. Includes variations Left, Right, Full, Outer, and Inner joins. \\ \hline
OrderBy          & Sorts the records of the table based on a specified column.                                                        \\ \hline
Aggregate & Performs a calculation on a set of values (records) and outputs a single value (record). Common aggregations include summation and multiplication. \\ \hline
GroupBy           & Groups the data using the given columns; GroupBy is usually followed by aggregate operations.                      \\ \hline
\end{tabular}
\end{table}

\begin{table}[htpb]
\caption{Operations as specified by BLAS} 
\label{tab:blass_op}
\centering
\begin{tabular}{|p{3cm}|p{13cm}|}
\hline
\textbf{Operation} & \textbf{Description}                                         \\ \hline
Level 1 & Operations on vectors i.e adding two vectors \\ \hline
Level 2 & Operations for combination of vectors and matrices. i.e. matrix and vector multiplication \\ \hline
Level 3 & Matrix operations i.e. matrix and matrix multiplication \\ \hline
\end{tabular}
\end{table}

\subsubsection{Distributed Operators}

A distributed operator works across data in multiple processes in many nodes of a cluster. A distributed operator needs communication options and local operators. Compared to the number of local operators defined on a data structure, there are a limited set of communication operators for a given data structure, and some of them are listed in Table~\ref{tab:array_op} where 720 MPI operators support classic parallel computing. Higher-level distributed operations are built by combining these communication operations with local operations, as shown in Table~\ref{tab:dist_op}. These include the famous MapReduce \citep{dean2008mapreduce} which abstraction showed clearly the similarity between distributed operators in the technical and database computing domains. MapReduce and its implementation in Hadoop enabled parallel databases as in Apache Hive. They added Group-By and key-value pairs to the Reduce operation common in the previous HPF family simulation applications. The powerful yet straightforward MapReduce operation was expanded in Big Data systems, primarily through the operators of Databases (union, join, etc.), Pandas, and the Spark, Flink, Twister2 family of systems. 
\newline

\begin{table}[htpb]
\caption{Communication operations for data structures} 
\label{tab:array_op}
\centering
\begin{tabular}{|p{3cm}|p{13cm}|}
\hline
\textbf{Data Structure} & \textbf{Operations}                                         \\ \hline
Arrays & Reduce, AllReduce, Gather, AllGather, Scatter, AllToAll, Broadcast, Point-to-Point \\ \hline
Tables & Shuffle (Similar to AllToAll but specifically designed for Tables), Broadcast \\ \hline
\end{tabular}
\end{table}

\begin{table}[htpb]
\caption{Higher level distributed operations} 
\label{tab:dist_op}
\centering
\begin{tabular}{|p{3cm}|p{13cm}|}
\hline
\textbf{Distributed Operation} & \textbf{Implementation}                                         \\ \hline
Sorting tables & Shuffle followed by a local sorting operation \\ \hline
Join tables & Partitioning of records, shuffle and local join operation \\ \hline
Matrix multiplication & Point to point communication and local multiplication \\ \hline
Vector addition & AllReduce with SUM \\ \hline
\end{tabular}
\end{table}

\subsection{Distributed Execution}

There are two main distributed execution methods used in current systems. They are asynchronous execution and loosely synchronous execution. In an asynchronous system, the parallel task instances can execute independently using message queues to decouple them in time. This is seen in systems like Spark and Hadoop. In a loosely synchronous system, the parallel tasks assume they can directly send messages to other similar jobs. It is called loosely synchronous because synchronization only happens when they need to communicate with each other. Otherwise, parallel task instances can work independently. 

~\archname{} architecture currently only supports the synchronous execution. The asynchronous execution demands the system to be tightly integrated with a central coordinator and a scheduler. This makes it harder to develop distributed operators independently and make them work together.

\section{\archname{} Frameworks}
\label{frameworks}
Now let us look at Cylon and Deep Learning frameworks and see how they can work together according to the \archname{} architecture. First, we describe how Cylon is designed to support distributed data engineering on a dataframe abstraction. Then we discuss how Cylon can be coupled with state-of-the-art deep learning frameworks to organise end-to-end data analytics workloads.   

\subsection{Cylon}

\begin{figure}[htpb]
\begin{center}
\includegraphics[width=0.45\textwidth]{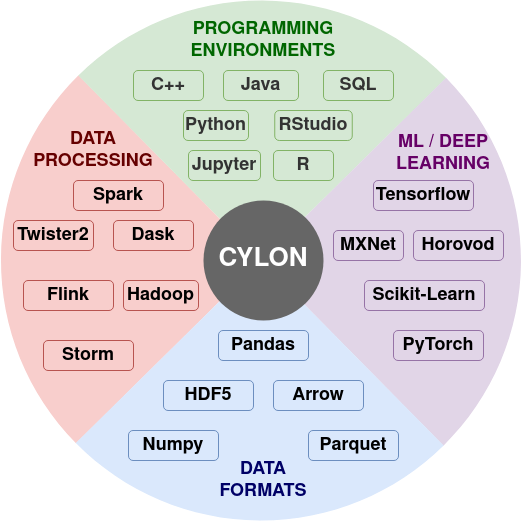}
\end{center}
\caption{Cylon for data engineering}
\label{fig:cylon-dataengineering}
\end{figure}

Cylon~\citep{widanage2020high, abeykoon2020data} provides a distributed memory DataFrame API on Python for processing data using a tabular format. Cylon provides a Python API around high-performance compute kernels in C++. These kernels are written on top of the Apache Arrow based efficient in-memory table representation. It can be deployed with MPI for distributed memory computations processing large datasets in HPC clusters. Operators in Cylon are based on relational algebra and closely resemble the operators in Pandas DataFrame to provide a consistent experience. The user can program with a global view of data by applying operations on them. Also, they can convert the data to local parallel processes and do in-memory operations as well. Cylon can be thought of as a framework that can work across different frameworks, data formats to connect various applications, as shown in Figure~\ref{fig:cylon-dataengineering}.

\begin{figure}[htpb]
\begin{center}
\includegraphics[width=0.45\textwidth]{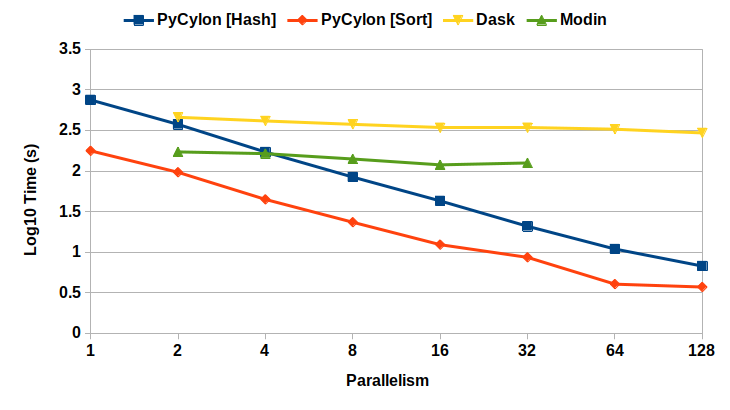}
\end{center}
\caption{Distributed Join Performance}
\label{fig:pycylon-distributed-join-performance}
\end{figure}

Cylon is different from other table abstractions such as Modin~\citep{petersohn2020towards}, Dask~\citep{rocklin2015dask} and Spark~\cite{spark2010} because it supports an efficient loosely synchronous execution model. These other frameworks use the asynchronous execution model, which relies on a central scheduler and a coordinator and does not conform to the ~\archname{} architecture. Figure~\ref{fig:pycylon-distributed-join-performance} shows how the Cylon Join operator performs compared to other frameworks. This experiment used 200M records per relation (for both left and right tables in a join) and scaled up to 128 processes. Random data were generated by considering the uniqueness of data to be 10\% such that the join performs under higher stress feeling hash functions and hash-based shuffles. In the parallel experiments, each process will be loading an equal amount of data such that the total amount is limited to 200M records. The results from Figure \ref{fig:pycylon-distributed-join-performance} show that our distributed join implementation is faster than Dask and Modin implementations. Also, the scalability in Dask and Modin is not very strong compared to the scaling provided by PyCylon. Also, the Modin couldn't be scaled up beyond a single machine and failed in the execution. 

\subsection{Deep Learning Frameworks}

Deep learning workloads are compute-intensive. Most of the existing deep learning frameworks can run codes in a distributed manner. Here, the widely used approach is the distributed data-parallel model. Distributed data-parallel model deals with the distributed memory architecture and has the loosely synchronous execution capability. 

PyTorch offers a distributed data-parallel (DDP) model, which allows the user to train large models using many GPUs. It can use distributed frameworks such as MPI, NCCL, or GLOO for the necessary communication operations for deep learning training with multiple GPUs. Tensorflow does loosely synchronous distributed execution via frameworks like Horovod. Due to these reasons, we can think of these systems as ~\archname{} when running data-parallel training using the loosely synchronous execution model. 
\newline

\subsection{Deep Learning \& Data Engineering}

Because the distributed execution of Cylon and deep learning systems such as PyTorch and Tensor conform to the ~\archname{} architecture, they can work together in a single parallel program. This improves productivity and usability in dealing with end-to-end analytical problems. In a data analytics-aware data engineering workload, three main factors govern usability and performance. 

\begin{itemize}
    \item Single source, including data engineering and data analytics
    \item Simple execution mode for sequential and distributed computing 
    \item Support for CPUs and GPUs for distributed execution
\end{itemize}

The single source refers to writing the data engineering and analytics code in a single script and executing with a single command. This is a beneficial and efficient method to do data exploration based data analytics. For such workloads, feature engineering and data engineering components are extensively modified to see how the data analytics workload performs for different settings. In such cases, the data scientist must have room to write the usual Python script and run the data analytics workload efficiently, not only in a single node but also across multiple nodes. Simple execution mode refers to running the workload with a simple method to spawn the processes to run in parallel. 

\begin{figure}[!h]
\begin{center}
\includegraphics[width=0.45\textwidth]{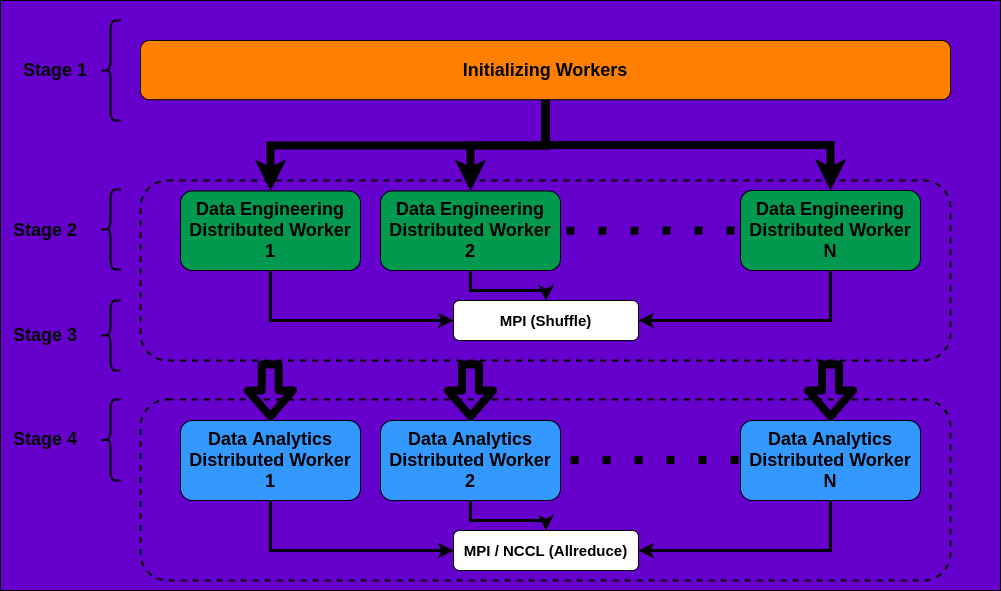}
\end{center}
\caption{Integrating Data Engineering Workload with Data Analytics Workload}
\label{fig:integrating-data-engineering-with-data-analytics-overview}
\end{figure}

Data analytics frameworks provide different methods to spawn parallel jobs. For instance, Dask requires that the user start the workers and schedulers on each node and provide host information for distributed communication. MPI allows for a single execution command \textit{mpirun} to spawn all the processes. Such factors are essential in providing a unified interface to do deep learning easily. Also, the execution mode on various accelerators for deep understanding is a vital component. The majority of the frameworks support both CPU and GPU execution, so it is essential to provide the means to seamlessly integrate with these execution models to support data analytics workloads. Figure \ref{fig:integrating-data-engineering-with-data-analytics-overview} highlights the high-level component overlay of a data analytics-aware data engineering workload. We have partitioned the workflow into four stages. 

\begin{itemize}
    \item Stage 1: In the first stage, the processes must be spawned depending on the parallelism. A unified process spawning mechanism that identifies worker information such as host IP addresses for each machine or network information is identified at this stage.
    \item Stage 2: Worker information is extracted, and data engineering operators will run in distributed mode on top of the data engineering platform, which depends on the worker initialization component. Here the operations can be distributed or pleasingly parallel.
    \item Stage 3: For data analytics workloads, the worker information, network information, chosen accelerator, and data must be provided from the corresponding data engineering process. This mapping is 1:1 for data engineering workers to data analytics workers. But this can also be a many-to-many relationship. 
    \item Stage 4: The worker information, network information and data will be used to execute the data analytics workload is distributed or pleasingly parallel mode. 
\end{itemize}

Considering this generic overview on deploying deep learning workloads with data engineering workloads, we have integrated PyCylon with distributed data-parallel models for PyTorch, Horovod-PyTorch, and Horovod-Tensorflow. Horovod is a distributed deep learning framework that supports a unified API for handling distributed deep learning on multiple frameworks. Horovod supports PyTorch, Tensorflow, and MXNet. In our research, we paid close attention to PyTorch and Tensorflow. Horovod internally uses mpirun to spawn the processes, and this model fits very well with PyCylon internals as we relied on mpirun to spawn the processes. This makes PyCylon uniquely qualified as a supportive data engineering framework for Horovod.

The first step is to initialize the runtime. Here either PyTorch distributed initialization, or PyCylon distributed initialization can be called. But especially on CPUs, the PyTorch initialization must be called since PyTorch internally does not handle the MPI initialization check. But if we use NCCL as the back-end, this constraint does not exist. This is one of the bugs we discovered from our previous research. For the PyTorch DDP, the master address and port must be provided because the NCCL back-end needs to identify which work will be designated as the master-worker to coordinate the communication. In addition, the initialization method has to be set. After the distributed initialization in PyTorch, the PyCylon context must be initialized to set to distributed mode. After this stage, we complete the requirements for stage 1 and partial requirements for stage 3 (network information is also passed along with data in stage 3, which is initialized in this step).  Figure \ref{fig:stage-1-intialization-for-pytorch-with-pycylon} is a sample code snippet related to the initialization step.

\begin{lstlisting}[language=Python, caption={Stage 1: Initialization for PyTorch With PyCylon}, label={fig:stage-1-intialization-for-pytorch-with-pycylon}]

os.environ['MASTER_ADDR'] = master_address
os.environ['MASTER_PORT'] = port
os.environ["LOCAL_RANK"] = str(rank)
os.environ["RANK"] = str(rank)
os.environ["WORLD_SIZE"] = str(world_size)
dist.init_process_group(backend=backend, init_method="env://")
mpi_config = MPIConfig()
env = CylonEnv(config=mpi_config, distributed=True)    
\end{lstlisting}

The data engineering workload is done in PyCylon, assuming the distributed mode initialization. We first join two tables and use the join response for a deep learning workload. The distributed join is called by providing the initialized context information to the join function. At the end of this stage, we create the resultant dataframe, and later on, in stage 3, this dataframe can be used to generate the Numpy array required for deep learning. This stage is typical for any framework, including PyTorch, Tensorflow, etc. 
Figure \ref{fig:stage-2-pycylon-distributed-data-engineering} details a sample data engineering workload for a data analytics problem.

\begin{lstlisting}[language=Python, caption={Stage 2: PyCylon Data Engineering Workload},label={fig:stage-2-pycylon-distributed-data-engineering}]
df1 = DataFrame(read_csv("...")) 
df2 = DataFrame(read_csv("...")) 

join_df = df1.merge(right=df2, left_on=[0], right_on=[3], algorithm='hash')
\end{lstlisting}

% \begin{figure}[htpb]
% \begin{center}
% \includegraphics[width=0.45\textwidth]{images/extension_frameworks/pytorch/stage_2_pycylon.png}
% \end{center}
% \caption{Stage 2: PyCylon Data Engineering Workload}
% \label{fig:stage-2-pycylon-distributed-data-engineering}
% \end{figure}

In stage 3, stage 2 is used to create tensors required for the deep learning stage. We also perform the data partitioning for training and testing. This stage is different from framework to framework since the tensor creation and data partitioning steps can have various internal utils. We do not use data loaders or data samplers but note that these tools can be used to generate both. Figure \ref{fig:stage-3-moving-data-from-data-engineering-to-data-analytics} is a sample code snippet for data movement from data engineering workload to data analytics workload. 

\begin{lstlisting}[language=Python, caption={Stage 3: Moving Data from Data Engineering Workload to Data Analytics Workload}, label={fig:stage-3-moving-data-from-data-engineering-to-data-analytics}]
data_ar: np.ndarray = feature_df.to_numpy()
df_ftrs: np.ndarray = data_ar[:, 0:3]
df_lrnr: np.ndarray = data_ar[:, 3:4]
x_train, y_train = df_ftrs[0:100], df_lrnr[0:100]
x_test, y_test = df_ftrs[100:], df_lrnr[100:]
...
x_train = torch.from_numpy(x_train).to(device)
y_train = torch.from_numpy(y_train).to(device)
x_test = torch.from_numpy(x_test).to(device)
y_test = torch.from_numpy(y_test).to(device)

\end{lstlisting}

% \begin{figure}[htpb]
% \begin{center}
% \includegraphics[width=0.45\textwidth]{images/extension_frameworks/pytorch/stage_3_pytorch_data.png}
% \end{center}
% \caption{Stage 3: Moving Data from Data Engineering Workload to Data Analytics Workload}
% \label{fig:stage-3-moving-data-from-data-engineering-to-data-analytics}
% \end{figure}

In stage 4, we initialize the deep learning model and the DDP model using the sequential model. We pass along device information such that tensors and models are copied to the corresponding devices (if accelerators are involved) for training and testing. This initialization part varies from framework to framework depending on the requirements and APIs. Figure \ref{fig:stage-4-distributed-data-analytics-workload} highlights the initialization of a DDP model with PyTorch.

\begin{lstlisting}[language=Python, caption={Stage 4: PyTorch Distributed Data Analytics Workload},label={fig:stage-4-distributed-data-analytics-workload}]
model = Network().to(device)
ddp_model = DDP(model, device_ids=[device])
loss_fn = nn.MSELoss()
optimizer = optim.SGD(ddp_model.parameters(), lr=0.01)
optimizer.zero_grad()
for t in range(epochs):
    for x_batch, y_batch in zip(x_train, y_train):
        prediction = ddp_model(x_batch)
        loss = loss_fn(prediction, y_batch)
        loss.backward()
        optimizer.step()
        optimizer.zero_grad()
\end{lstlisting}

\subsubsection{Horovod with PyTorch}

Horovod PyTorch provides the ability to scale on both GPUs and CPUs with a unified API. This is significant because PyTorch does not need to be compiled from the source to get MPI capability. Horovod has already offloaded the distributed trainer, optimizer, and allreduce communication packages. The internal DDP mechanism that does this in PyTorch is offloaded. 

In stage 1, the Horovod init method must be called to initialize the environment. After that, the Cylon context can be initialized with distributed runtime true. If GPUs are used, the correct device must be set to PyTorch CUDA configs. To obtain the device IDs, we can either use the rank from Horovod initialization or PyCylon initialization. Still, at the moment, Horovod supports local rank as well, and it is more suitable in terms of effortlessly integrating with the distributed runtime for Horovod-PyTorch. Figure \ref{fig:stage-1-intialization-for-pytorch-horovod-with-pycylon} shows a sample code snippet demonstrating how this is accomplished.

\begin{lstlisting}[language=Python, caption={Stage 1: Initialization for Horovod-PyTorch With PyCylon},label={fig:stage-1-intialization-for-pytorch-horovod-with-pycylon}]
hvd.init()
mpi_config = MPIConfig()
env = CylonEnv(config=mpi_config, distributed=True)
rank = env.rank
cuda_available = torch.cuda.is_available()
device = 'cuda:' + str(rank) if cuda_available else 'cpu'
if cuda_available:
    # Horovod: pin GPU to local rank.
    torch.cuda.set_device(hvd.local_rank())
    torch.cuda.manual_seed(42)
\end{lstlisting}

Another essential thing to note is that the data engineering code remains the same for any deep learning framework discussed in this context. Also, as with the PyTorch data engineering section, the output can be converted to a Numpy array using the endpoints from the PyCylon dataframe. Also, the tensors can be created by providing the device IDs obtained from the Horovod runtime, and data can be prepared for a deep learning workload. 

In stage 4, following the tensor creation step, the Horovod-related initialization must be done to prepare the optimizers, network and other utils for distributed training. PyTorch-Horovod integration, PyTorch's default neural network model, loss function, and optimizer can be used as input to the distributed computation-enabled Horovod components. First, the model parameters and optimizer must be broadcast using the Horovod broadcast method from $0^{th}$ rank. There are two method calls designated for initial network values and optimizer values. Also, Horovod provides a compression algorithm to select whether compression is required for distributed communication. After these steps, the distributed optimizer must be set by passing the initialized values. Figure \ref{fig:stage-4-distributed-data-analytics-workload-pytorch-horovod} includes a sample code snippet to initialize the Horovod components for distributed data-parallel deep learning with PyTorch.

\begin{lstlisting}[language=Python, caption={Stage 4: Distributed Data Analytics PyTorch-Horovod Workload},label={fig:stage-4-distributed-data-analytics-workload-pytorch-horovod}]
optimizer = optim.SGD(...)
hvd.broadcast_parameters(model.state_dict(), 
                        root_rank=0)
hvd.broadcast_optimizer_state(optimizer, root_rank=0)
compression = hvd.Compression.fp16
model_ps = model.named_parameters()
optimizer = hvd.DistributedOptimizer(optimizer, named_parameters=model_ps,
                    compression=compression, op=hvd.Adasum,
                    gradient_predivide_factor=1.0)
\end{lstlisting}

\subsubsection{Horovod with Tensorflow}

Similar to PyTorch integration, Horovod also supports Tensorflow. Tensorflow has its own distributed training platform. It contains distributed mirrored strategy as the equivalent routine for distributed data-parallel training. To start this run, we initialize Horovod and PyCylon. As with PyTorch, we also need to decide how the device is selected depending on the accelerator. The Tensorflow config API provides a listing of GPUs, and this information is added to the Tensorflow configurations to make all the GPU devices available. Figure \ref{fig:stage-1-intialization-for-horovod-tensorflow-with-pycylon} is a code snippet for the aforementioned initialization. 

\begin{lstlisting}[language=Python,caption={Stage 1: Initialization for Tensorflow With PyCylon},label={fig:stage-1-intialization-for-horovod-tensorflow-with-pycylon}]
hvd.init()
assert hvd.mpi_threads_supported()
mpi_config = MPIConfig()
env = CylonEnv(config=mpi_config, distributed=True)
rank = env.rank
world_size = env.world_size
gpus = tf.config.experimental.list_physical_devices('GPU')
for gpu in gpus:
    tf.config.experimental.set_memory_growth(gpu, True)
if gpus:
    tf.config.experimental.set_visible_devices(gpus[hvd.local_rank()], 'GPU')
\end{lstlisting}

Similar to prior experience, the data engineering component also remains unchanged for Horovod-Tensorflow integration. The data analytics data structure creation is different from framework to framework. Tensorflow has its own set of APIs to make these steps simpler and more structured. The Tensorflow dataset API can be used to create tensors from Numpy arrays, and this API can be used to shuffle and create mini-batches, as expected by the deep learning workload. Figure \ref{fig:stage-3-moving-data-from-data-engineering-to-data-analytics-horovod-tensorflow} contains a code snippet detailing this step. 

\begin{lstlisting}[language=Python,caption={Stage 3: Moving Data from Data Engineering Workload to Data Analytics Workload},label={fig:stage-3-moving-data-from-data-engineering-to-data-analytics-horovod-tensorflow}]
...
train_dataset = tf.data.Dataset.from_tensor_slices((x_train, y_train))
test_dataset = tf.data.Dataset.from_tensor_slices((x_test, y_test))
BATCH_SIZE = 64
SHUFFLE_BUFFER_SIZE = 100
train_dataset = train_dataset.shuffle(SHUFFLE_BUFFER_SIZE).batch(BATCH_SIZE)
test_dataset = test_dataset.batch(BATCH_SIZE)
...
\end{lstlisting}

Horovod-Tensorflow also requires a set of initialization steps to train a Tensorflow deep learning model. Like PyTorch, the Tensorflow loss function, optimization function and neural network model are compatible with Tensorflow-Horovod internals. The gradient tape from Tensorflow autograd can be used, and for this, Horovod provides a DistributedGradientTape operator, which takes the gradient tape instance as a parameter. In addition, before training, this DistributedGradientTape must be initialized with the model parameters and loss function, and the optimizer values must be set to initial values. Again, the model parameters and optimizer values must be broadcast using designated Horovod broadcast functions. Figure \ref{fig:stage-4-distributed-data-analytics-horovod-tf-workload} illustrates this. 

\begin{lstlisting}[language=Python,caption={Stage 4: Distributed Data Analytics Horovod-Tensorflow Workload},label={fig:stage-4-distributed-data-analytics-horovod-tf-workload}]
model = tf.keras.Sequential(...)
loss = tf.losses.MeanSquaredError()
opt = tf.optimizers.Adam(0.001 * hvd.size())

@tf.function
def training_step(images, labels, first_batch):  
    with tf.GradientTape() as tape:
        probs = model(images, training=True)
        loss_value = loss(labels, probs)
    tape = hvd.DistributedGradientTape(tape)
    grads = tape.gradient(loss_value, model.trainable_variables)
    opt.apply_gradients(zip(grads, model.trainable_variables))
    if first_batch:
        hvd.broadcast_variables(model.variables, root_rank=0)
        hvd.broadcast_variables(opt.variables(), root_rank=0)
    return loss_value
\end{lstlisting}

\section{UNOMT Application}
\label{unomt}
% \label{s:unomt-application}

%%%% TODOs: UNOMT Application
%% 1. Need to migrate images from the thesis [IN PROGRESS]
%% 2. Content placement [TODO]
%% 3. Content re-write or rephrase.[TO DO]
%% 4. Review 1 [TO DO]
%% 5. Review 2 [TO DO]
%% Discuss how operator based approach is used to make UNO parallel application. [Summary Point at the end or beginning]

To demonstrate an end-to-end \archname{} architecture, we implemented a scientific application with a workload containing data engineering and data science computations. Our objective is to showcase how a sequential workload can be designed in a distributed manner using PyCylon and run a deep learning workload seamlessly on only a single script with a unified runtime. For this, we selected an application that uses Pandas dataframe for data engineering and PyTorch for data analytics. The original application is sequentially executed, and we have implemented a parallel version of this application with PyCylon and distributed PyTorch.
\newline

\subsection{Background}

% \label{s:uno}

UNOMT application is part of CANDLE\cite{wozniakhigh,xia2021cross} research conducted by Argonne National Laboratory, focusing on automated detection of tumour cells using a deep learning approach. The uniqueness of this approach is the composition of a data engineering workload followed by a deep learning workload written in PyTorch. This provides an ideal scientific experiment to showcase multiple systems working together to facilitate an efficient data pipeline. The goal of the UNOMT application is to give a cross-comparison of cancer studies and integrate it into a unified drug response model. Cell RNA sequences, drug descriptors and drug fingerprints are used as such responses to train the model. 

In the deep learning component, multiple networks are involved working on small and large datasets in the training process. Our research focuses on the more extensive network designed to calculate the drug response based on the cell-line information.
\newline

\subsection{Deep Learning Component}

UNOMT refers to a unified deep learning model with multi-tasks to predict drug response as a function of tumour and drug features for personalized cancer treatment. Precision oncology focuses on providing medicines for specific characteristics of a patient's tumour. The drug sensitivity is quantified by drug dose-response values which measure the ratio of treated to untreated cells after treatment with a specific drug concentration. In this application, a set of drug data obtained from the NCI60 human tumour cell line database\cite{shoemaker2006nci60} is used to predict the drug response by considering gene expression, protein and microRNA abundance. As per the contemplated scope, the UNOMT application we focus on in the study is conducted on single-drug response prediction using NCI60 and gCSI datasets. We used 1006 drugs from NCI60 database for this evaluation and gCSI for the cross-validation. The original application runs sequentially, and our contribution is providing a parallelized runtime for data engineering and running the deep learning workload alongside it. 

% To evaluate the drug response predictions (regression model), the metrics selected are $R^{2}$ (explained variance) and mean absolute error (MAE). The input features used to evaluate drug response are the cell-line gene expression profiles, drug chemical descriptors and molecular fingerprints. Here the drug response is modelled as a function of cell-line features and drug properties. The input features are engineered such that RNAseq expression profiles, drug descriptions, drug fingerprints and drug concentration are used as input parameters for the deep learning model. 

% Drug response regression network is an ensemble model which uses two other networks to support the classification. Figure \ref{fig:uno-dnn-architecture-drug-network} refers to the drug network, which also has 3 dense layers. This network is pre-trained before being used in the drug response regression network. 

% \begin{figure}[htpb]
% \begin{center}
% \includegraphics[width=0.20\textwidth]{images/uno_app/gene_network.png}
% \end{center}
% \caption{UNO DNN Architecture: Gene Network}
% \label{fig:uno-dnn-architecture-gene-network}
% \end{figure}

% \begin{figure}[htpb]
% \begin{center}
% \includegraphics[width=0.20\textwidth]{images/uno_app/drug_network.png}
% \end{center}
% \caption{UNO DNN Architecture: Drug Network}
% \label{fig:uno-dnn-architecture-drug-network}
% \end{figure}

% \begin{figure}[htpb]
% \begin{center}
% \includegraphics[width=0.20\textwidth]{images/uno_app/resp_block_net.png}
% \end{center}
% \caption{Response Block Module}
% \label{fig:uno-dnn-architecture-resp-block-module}
% \end{figure}

\begin{figure}
  \centering
  \includegraphics[width=0.15\textwidth]{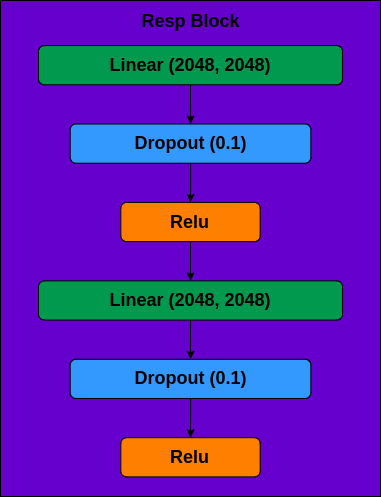}
  \caption{Response Block Module}
  \label{fig:uno-dnn-architecture-resp-block-module}
\end{figure}

The drug response model contains a dense input layer of shape 1537 to get the concatenated results of the gene network and the drug network response along with the concentration value. Within the drug response regression network, there is another residual block being used repeatedly. This layer is called the drug response block module, which contains two dense layers followed by a dropout layer and a ReLU activation layer. Figure \ref{fig:uno-dnn-architecture-resp-block-module} depicts the response block module. 

\begin{figure}[htpb]
\begin{center}
\includegraphics[width=0.95\textwidth]{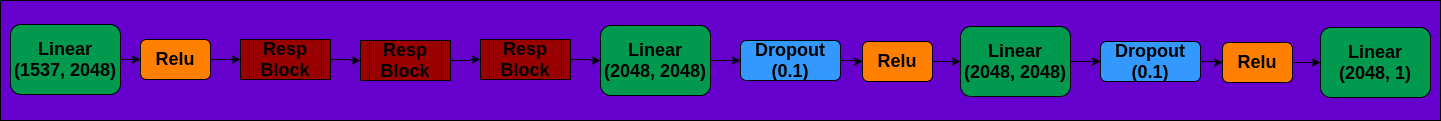}
\end{center}
\caption{Response Network}
\label{fig:uno-dnn-architecture-response-network}
\end{figure}

Residual blocks are stacked, and a set of dense layers are as well. Finally, the regression layer contains a single output dense layer. The number of response blocks can be customized dynamically and the number of dense layers that follow it. All these parameters can be provided as a hyper-parameter in the application configuration file. Figure \ref{fig:uno-dnn-architecture-response-network} shows the drug response regression network. 

This network is trained in a distributed data-parallel model since it contains a vast dataset and a complex network compared to the other examples introduced simultaneously. The corresponding data engineering component is also distributed data-parallel, which is discussed in detail in Section \ref{s:s:data-engineering-component}.

\subsection{Data Engineering Component}\label{s:s:data-engineering-component}

UNOMT application uses 2.5 million samples of cancer data across six research centres. This model analyses the study bias across these samples to design a unified drug response model. Before building this model, the application consists of a data engineering workload written in Pandas. The application consists of a few data engineering operators: concat (inner-join), to\_csv, rename, read\_csv, astype, set\_index, map, isnull, drop, filter, add\_prefix, reset\_index, drop\_duplicates, not\_null, isin and dropna.

The existing data engineering workload is written in Pandas and does not run in parallel. We re-engineered this application to a parallel data engineering workload. We designed a seamless integration between data analysis and data engineering workload consuming state-of-the-art high-performance computing resources. We also integrated a Modin-based implementation to showcase the performance comparison with our implementation. The data engineering workload is executed in CPU-based distributed memory, and the data analytical workload can be either run in CPU or GPU. We use Pytorch for data analytics workload and extend it to PyTorch distributed data-parallel training. Our objective is to integrate an HPC-based full stack of data analytics-aware data engineering for scalability. PyCylon only supports this feature at the moment. Also, we stress the importance of designing a BSP-based model for deep learning workloads associated with data engineering components for better performance and scalability in HPC hardware.

\begin{figure*}
\centering
\begin{minipage}{.3\textwidth}
  \centering
  \includegraphics[width=0.45\textwidth]{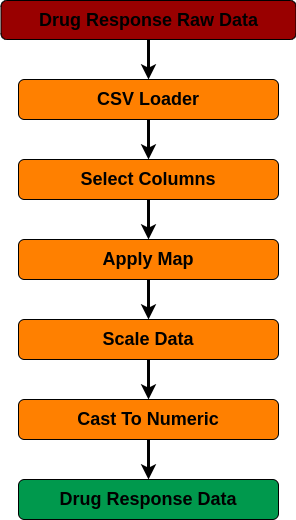}
  \caption{Drug Response Data Processing}
  \label{fig:uno-drug-response-data-processing}
\end{minipage}%
\begin{minipage}{.3\textwidth}
  \centering
  \includegraphics[width=0.75\textwidth]{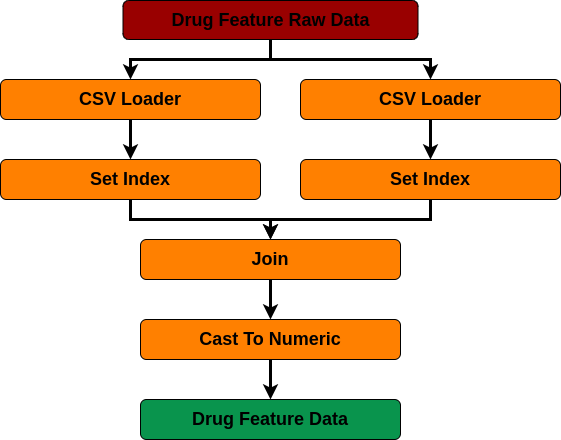}
  \caption{Drug Feature Data Processing}
  \label{fig:uno-drug-feature-data-processing}
\end{minipage}
\begin{minipage}{.3\textwidth}
  \centering
  \includegraphics[width=0.45\textwidth]{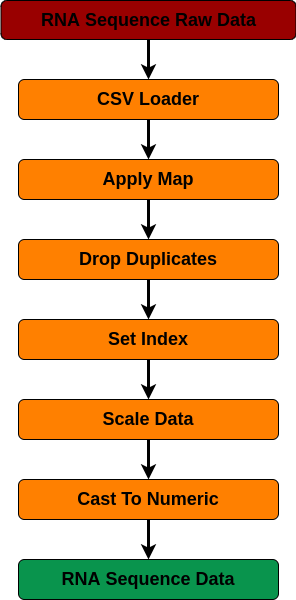}
  \caption{RNA Sequence Data Processing}
  \label{fig:uno-rna-sequence-data-processing}
\end{minipage}
\end{figure*}

The data analytics component requires a set of features to be engineered from the raw data. Here, three primary datasets are necessary to create the complete dataset for the drug response model. Figure \ref{fig:uno-drug-response-data-processing} refers to the primary dataset, which contains the drug response. The raw dataset possesses additional features, so the data is loaded in the initial stage, and a column filtering operation selects extract the expected features. Then a map operation is performed to preprocess a drug ID column to remove symbols from the columns and create a consistent drug ID. Once the data are cleaned, they are scaled with the Scikit-learn preprocessing library for scaling numerical values. After this, the data are fully converted into a numeric type to provide numeric tensors for the deep learning workload. In the parallel mode, we partition this dataset with the set parallelism, upon which it is passed to the corresponding operators. 

% \begin{figure}[htpb]
% \begin{center}
% \includegraphics[width=0.15\textwidth]{images/uno_app/drug_response_raw_data_processing.png}
% \end{center}
% \caption{Drug Response Data Processing}
% \label{fig:uno-drug-response-data-processing}
% \end{figure}

To formulate the global dataset, we require two other datasets which act as metadata to filter and process the primary drug response dataset. The first is the drug feature raw dataset, which contains drug features required to be located in the drug response data. Two sub-datasets contribute to formulating the drug feature dataset. We merge them by performing an inner join on the dataset based on the index formed on the drug IDs. After that, we cast the data into numeric types and output them as a numeric array which is later converted to a numeric tensor for deep learning. This data processing workflow is shown in Figure \ref{fig:uno-drug-feature-data-processing}.

% \begin{figure}[htpb]
% \begin{center}
% \includegraphics[width=0.25\textwidth]{images/uno_app/drug_feature_data_engineering.png}
% \end{center}
% \caption{Drug Feature Data Processing}
% \label{fig:uno-drug-feature-data-processing}
% \end{figure}

The other dataset required is the RNA sequence dataset containing information about RNA sequences. Here the dataset is first processed to remove specific symbols by a map operation, and then duplicate records are dropped by a drop duplicate operator. Then an index is set for this dataset, and later on, scaling is done on the numeric data using the Scikit-Learn preprocessing library. Finally, the data is cast to a numeric type, and preprocessed RNA-sequence data are formulated as a Numpy array, which is later converted into a numeric tensor for the deep learning workload. This data processing pipeline can be found in Figure \ref{fig:uno-rna-sequence-data-processing}.

% \begin{figure}[htpb]
% \begin{center}
% \includegraphics[width=0.15\textwidth]{images/uno_app/rna_sequence_data_engineering.png}
% \end{center}
% \caption{RNA Sequence Data Processing}
% \label{fig:uno-rna-sequence-data-processing}
% \end{figure}

\begin{figure}[htpb]
\begin{center}
\includegraphics[width=0.35\textwidth]{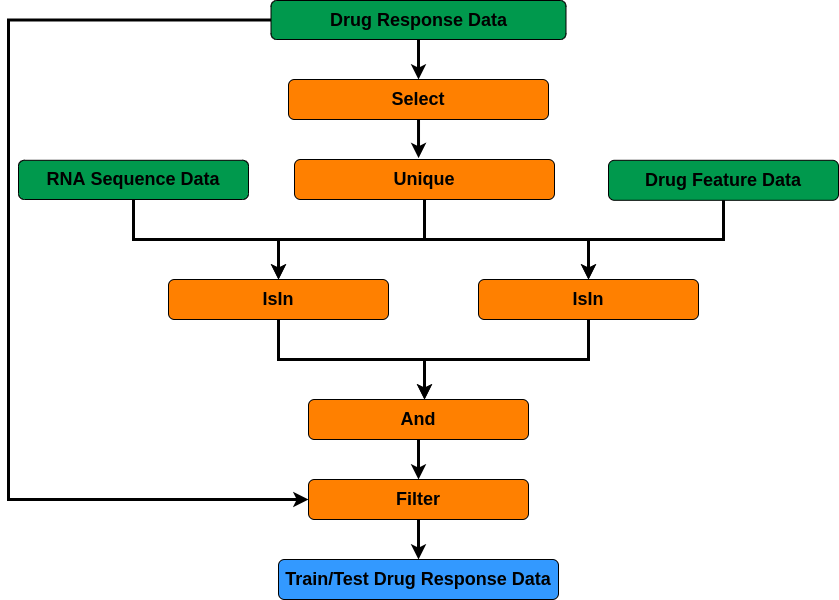}
\end{center}
\caption{Drug Response Overall Data Processing}
\label{fig:uno-drug-response-overall-data-processing}
\end{figure}

Once the drug response initial dataset, drug feature data and RNA-sequence data are preprocessed, the final dataset for the drug response model is engineered as shown in Figure \ref{fig:uno-drug-response-overall-data-processing}. The processed drug response data are further feature-selected, and a unique operation is applied. Then the RNA sequence data is filtered by checking whether specific drug-related RNA sequences are present. The same is done for the drug feature dataset. These two operations are done by the isin operator. Afterwards, the common drug set is selected by performing an and operation, and later these common drug-related drug response data filters are used to get the final drug response data. 

Among the operators applied, since we partitioned the data, each data engineering operator can work independently in a pleasingly parallel manner. But we can rely on the distributed unique operator to ensure no duplicate records are used for deep learning across all processes. Note that the data engineering component of this application is feature engineering metadata, and we use them to filter an extensive dataset converted to formulate the expected input for the drug response model. 

\section{Performance Evaluation}
\label{s:s:performance-evaluation}

The original application was a single-threaded application implemented on Pandas for data engineering and PyTorch for deep learning. Our first goal was to implement the sequential version of the application and improve the sequential performance. After the first stage, we conducted distributed experiments to see how we could scale our workload on CPUs for data engineering. We also extended the deep learning component of this application by integrating with PyTorch distributed execution framework on both CPUs and GPUs using MPI and NCCL, respectively. Our goal was to seamlessly incorporate a deep learning-aware data engineering workload using a single Python data engineering and deep learning script with a single runtime in this benchmark. Also, note that we used the drug response network-related more extensive data distribution for the application benchmark. At the same time, the smaller networks require a much shorter execution time than this larger model. 

For the experiments, we had two sets of clusters for CPUs and GPUs. Victor cluster of Future Systems was used with six nodes and 16 processes per each on the maximum parallelism for CPUs. This cluster contains Intel(R) Xeon(R) Platinum 8160 CPU @ 2.10GHz machine per node. GPU experiments had Tesla K80s with 8 GPU devices on Google Cloud Platform. For single-node single-process executions, we used the same Victor nodes. Pandas, PyCylon (single-core) and Modin (single-core) were deployed for the sequential performance comparisons. Finally, for the distributed performance comparisons, we used PyCylon and Modin on single node multi-core scaling. We selected Modin instead of Dask because it is closer to the data engineering stack proposed by PyCylon due to eager execution and the ability to convert an existing Pandas data engineering workload in a straightforward manner.

\begin{figure}[htpb]
\begin{center}
\includegraphics[width=0.45\textwidth]{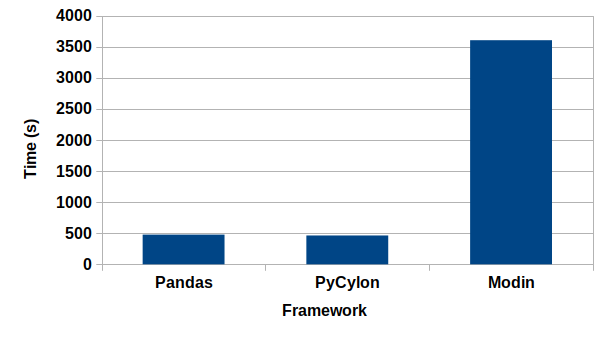}
\end{center}
\caption{Sequential Data Engineering}
\label{fig:uno-sequential-de-cpu}
\end{figure}

We first conducted experiments to evaluate the proposed systems' single process execution, PyCylon, Modin and Pandas. Modin provides the ability to convert a Pandas data engineering workload utilizing a single line of code. In contrast, PyCylon offers a dynamic API allowing the user to dynamically decide the nature of sequential and parallel operators. We evaluated the data engineering performance for the drug response data preprocessing workload used for the drug response regression network. Figure \ref{fig:uno-sequential-de-cpu} has the single-core performance for the aforementioned data engineering workload. We observe that the version of PyCylon and Pandas are very similar, while Modin is much slower. This performance improvement includes data loading efficiency plus overall operator performance improvements. But in a general way, Pandas and PyCylon have almost similar performance in most operators except for data loading, duplicate handling, null handling and search operations involved in this application. Note that both PyCylon and Modin are evolving data engineering frameworks to support data engineering on tabular data. 

% In the distributed performance evaluation Section \ref{s:s:data-engineering-distributed-performance}.

There are some data engineering components in this application like data normalization, parameterized data partition, and statistical data processing with third-party Python libraries like Scikit-Learn data processing and other statistical libraries. These libraries are compatible with Pandas data structure very well. Since PyCylon is seamlessly integrated for conversion to Pandas and back-and-forth, such third-party libraries can be easily used without any performance degradation. But for Modin, it cannot go back-and-forth between the Pandas data structure. This caused some of these operations to be relatively slower for Modin, compared to Pandas and PyCylon. This shows that we have to go beyond the dataframe construct and integrate with third-party libraries in implementing real-world applications. And to integrate with such libraries, data engineering frameworks must be very well designed with widely used data structures used by data scientists. 

\begin{figure}[htpb]
\begin{center}
\includegraphics[width=0.45\textwidth]{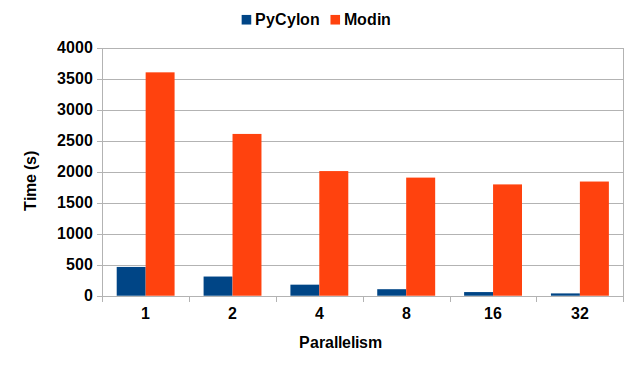}
\end{center}
\caption{Multi-Core Data Parallel Data Engineering Performance}
\label{fig:uno-multi-core-modin-vs-pycylon-de-tb-cpu}
\end{figure}

Since we scale this application by running in the distributed data-parallel setting, we compared the parallel performance for a single-node multi-core execution. Figure \ref{fig:uno-multi-core-modin-vs-pycylon-de-tb-cpu} shows the results for that application. These results show that the PyCylon is scaling well compared to Modin in the distributed data-data parallel setting. 

\begin{figure}[htpb]
\begin{center}
\includegraphics[width=0.45\textwidth]{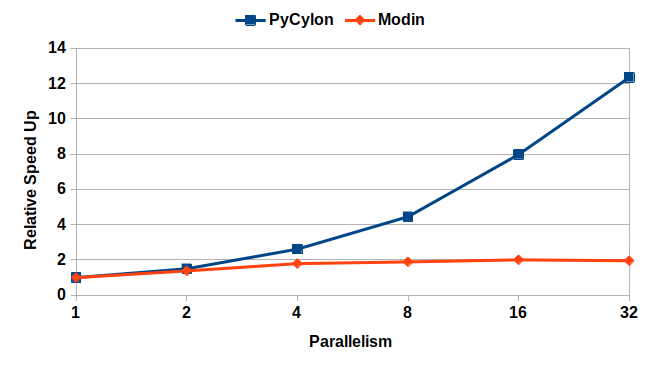}
\end{center}
\caption{Multi-Core Data Parallel Data Engineering Speed-up}
\label{fig:uno-multi-core-modin-vs-pycylon-de-speed-up}
\end{figure}

To understand the impact of PyCylon clearly, we also see that the relative speed-up for each framework is scoring a better value for PyCylon compared to Modin as shown in figure \ref{fig:uno-multi-core-modin-vs-pycylon-de-speed-up}. We observed the same scaling when we compare the performance for distributed-join operation. The main observation drawn from such behaviour is that the BSP execution model suits the data engineering applications well. 

\begin{figure}[htpb]
\begin{center}
\includegraphics[width=0.45\textwidth]{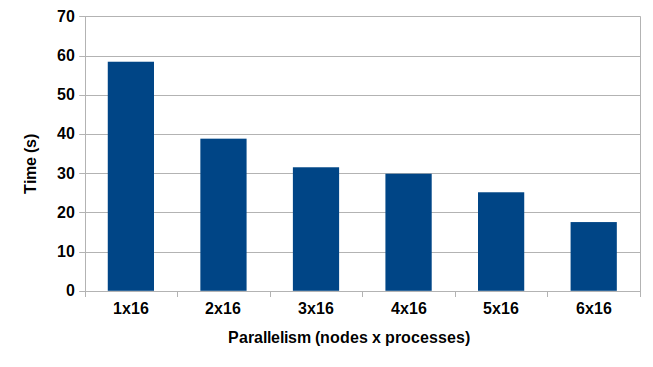}
\end{center}
\caption{PyCylon Distributed Data Parallel Data Engineering}
\label{fig:uno-distributed-de-cpu}
\end{figure}

We extended the distributed experiments further for multi-node multi-core. We observed that Modin failed to scale beyond a single node and failed in the cluster set-up. This could be a lack of documentation or an issue with the distributed framework Modin uses. Modin doesn't contain its own distributed runtime but relies on Ray or Dask. But with PyCylon, the only additional requirement is a host file and mapping of cores. Once that is provided, PyCylon can run the application end-to-end. The distributed data engineering performance for PyCylon is shown in figure \ref{fig:uno-distributed-de-cpu}. 

The deep learning experiments also extend data engineering runs on CPUs but both CPU and GPUs. In this context of investigations, we limited the experiment configuration such that, number of processes involved for data engineering and deep learning is the same. But PyCylon can be further improved to run in many-to-many process mapping for more complex data-parallel executions.

\begin{figure}[htpb]
\begin{center}
\includegraphics[width=0.45\textwidth]{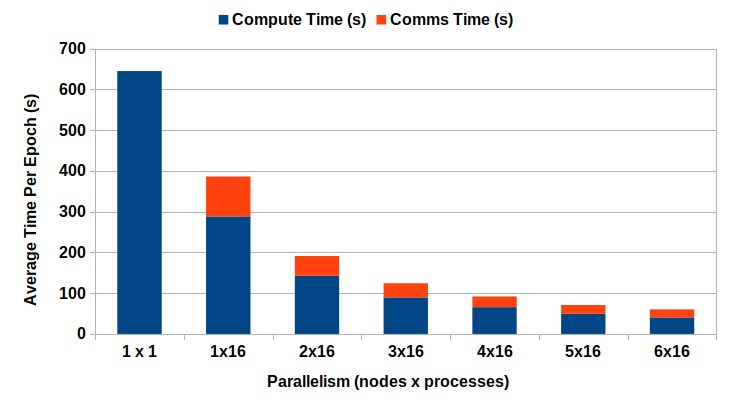}
\end{center}
\caption{Distributed Data Parallel Deep Learning on CPU}
\label{fig:uno-distributed-da-cpu}
\end{figure}

We selected PyTorch distributed communication framework with MPI for CPUs and NCCL for GPUs for the data analytics scaling experiments. The single process experiment results are the same for PyCylon and Pandas, and both have the same PyTorch codebase. Furthermore, all the data were in memory before the deep learning workload, so there was no overhead in loading data to create mini-batches. The experiments conducted on CPUs scaled well across multi-nodes, but we observed a slight memory overhead, causing the application to scale below the ideal point. We completed more experiments to evaluate an overhead from the data engineering framework, but we observed no significant overheads causing less scaling on CPUs. Figure \ref{fig:uno-distributed-da-cpu} highlighted the single process and distributed experiments carried out on CPUs. We used PyTorch built from the source to enable MPI execution, as it is a requirement forced by the framework. One significant factor is that PyTorch becomes an ideal distributed computation deep learning framework for PyCylon since PyCylon also supports an MPI backend for distributed computation. 

\begin{figure}[htpb]
\begin{center}
\includegraphics[width=0.45\textwidth]{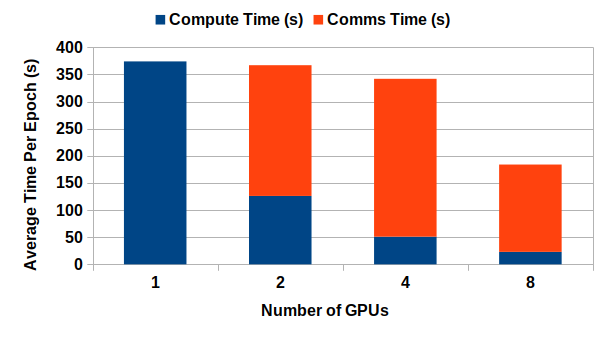}
\end{center}
\caption{Distributed Data Parallel Deep Learning on GPU}
\label{fig:uno-distributed-da-gpu}
\end{figure}

The GPU-based experiments were handled with a single-node multi-GPU experiment setting to see how the data analytics workload could be scaled on the NCCL execution framework with PyTorch. Figure \ref{fig:uno-distributed-da-gpu} displays the results for single GPU and multi-GPU experiments. We observed that the execution time was dominated by the communication time. With the increase of parallelism, the number of communications across devices increases, but the number of batches that has to be sent across devices decreases. This gives an advantage in scaling. When we consider the computation time, we saw that scaling happens closer to the ideal scaling point in all parallel settings. In addition, the computation is much faster in Parallelism 2 than in Parallelism 1, where the memory overhead is 50\% less than the sequential execution. When considering CPU vs GPU performance for the deep learning workload, the speed-up from GPUs is 2x compared to CPUs in this network.

\section{Related Work}
\label{related}

There are many efforts to build efficient distributed operators for data science and data engineering. Frameworks like Apache Spark~\citep{apache-spark}, Apache Flink~ \citep{carbone2015apache} and Map Reduce~\citep{dean2008mapreduce} are legacy systems created for data engineering. And many programming models have been developed on top of these big data systems to facilitate data analysis\citep{doi:10.1080/17445760.2017.1422501}. Later on, these systems adopted the data analytics domain under their umbrella of big data problems. But with the emerging requirement for high-performance computing for data science and data engineering, the existing parallel operators in these frameworks don't provide adequate performance or flexibility\citep{elshawi2018big}. Frameworks like Pandas \cite{pandas} gained more popularity in the data science community because of their usability. Pandas only provide serial execution, and Dask~\citep{rocklin2015dask} uses it internally (parallel Pandas) to provide parallel operators. Also, it was re-engineered as Modin~\citep{petersohn2020towards} to run the dataframe operators in parallel. But these efforts are mainly focused on a driver-based asynchronous execution model, a well-known bottleneck for distributed applications. 

The majority of the data analytics workloads tend to use data-parallel execution or bulk synchronous parallel (loosely synchronous) mode. Frameworks like PyTorch~\citep{pytorch} adopted this HPC philosophy, and distributed runtimes like Horovod~\citep{sergeev2018horovod} generalized this practice for most of the existing deep learning frameworks. They were adopting this philosophy along the same time HPC-driven big data systems like Twister2~\citep{twister2, wickramasinghe2019twister2, abeykoon2019streaming} were created to bridge the gap between data engineering and deep learning. But with the language boundaries of Java~\citep{ekanayake2016java} and usability with native-C++ based Python implementations were favoured over JVM-based systems. PyCylon~\citep{abeykoon2020data} dataframes for distributed CPU computation and Cudf~\citep{hernandez2020performance} dataframes for distributed GPU computation were designed. The seamless integration of data engineering and deep learning was a possibility with such frameworks and nowadays are being widely used in the data science and data engineering sphere to do rapid prototyping and design production-friendly applications.  

\section{Conclusions}
\label{conclude}

We showcased an example of the \archname{} architecture in an en end-to-end application where data engineering and deep learning operators working together in a single distributed program. This highlighted the importance of \archname{} based distributed and local operators on a different data structure that can work together in a single program. Further, the \archname{} style operators are more efficient in executing at scale, due to their loosely synchronous nature. Dask and Spark use an asynchronous execution model that requires a central server to schedule and coordinate the execution that can slow down applications at scale. Such asynchronous execution is not supported in the \archname{} because it is harder for different distributed operators to work together due to the dependency on the schedulers.  

\section*{Acknowledgments}
This work is partially supported by the National Science Foundation (NSF) through awards CIF21 DIBBS 1443054, SciDatBench 2038007, CINES 1835598 and Global Pervasive Computational Epidemiology 1918626. We thank the FutureSystems team for their infrastructure support. 

\bibliographystyle{frontiersinSCNS_ENG_HUMS}
\bibliography{ref.bib}

\end{document}